\documentclass[cernpreprint, atlasdraft=false, UKenglish, texmf, orcidlogo]{atlasdoc}
\usepackage{atlaspackage}
\usepackage{atlasbiblatex}

\usepackage{atlasphysics}
\graphicspath{{logos/}{figures/}}

\usepackage{ANA-BPHY-2022-03-PAPER-defs}

\AtlasTitle{Observation of a $B_c^{*+}$ meson with the ATLAS detector}

\AtlasAbstract{%
The first observation of a new meson state decaying into $B_c^+$ and a photon is presented
using a dataset of $pp$ collisions at a centre-of-mass energy $\sqrt{s} = 13$\,\TeV
collected by the ATLAS detector at the Large Hadron Collider,
corresponding to an integrated luminosity of 140\,fb$^{-1}$.
The $B_c^+$ mesons are reconstructed from a three-muon final state
from $B_c^+\to J/\psi(\mu^+\mu^-)\mu^+\nu_\mu X$ decays,
along with photons reconstructed via conversions to electron--positron pairs in the detector material.
The new state is observed with a significance that exceeds 8 standard deviations.
The mass difference between the new meson state and the ground-state $B_c^+$ meson
is measured to be
$64.5 \pm 1.4\stat{}^{+1.0}_{-1.4}\syst$\,\MeV.
This corresponds to a mass for the observed state of
$6339.0 \pm 1.4\stat{}^{+1.0}_{-1.4}\syst\pm 0.3\errmbc$\,\MeV, where the last uncertainty
is due to the precision of the best knowledge of the $B_c^+$ meson mass.
The low mass difference value
matches the theory expectations for the lowest vector state of the $\bar{b}c$ system,
the $B_c^{*+}$ meson.
}

\AtlasRefCode{BPHY-2022-03}

\PreprintIdNumber{CERN-EP-2026-142}

\AtlasJournal{PRL}
\AtlasJournalRef{\PRL 137 (2026) 141901}
\AtlasDOI{10.1103/vrxp-zkbs}

\AtlasCoverEgroupAnalysisTeam{atlas-bphy-2022-03-analysis-team@cern.ch}

\hypersetup{pdftitle={ATLAS document},pdfauthor={The ATLAS Collaboration}}

\begin{document}

\maketitle

The \Bcp meson represents the ground state of a family of mesons consisting of a $b$-antiquark and a $c$-quark.
These states share some properties of heavy quarkonium systems,
such as charmonium (\ccq) and bottomonium (\bbq),
and similar approaches are often used to theoretically study their mass spectrum~\cite{QuarkoniumWorkingGroup:2004kpm}.
However, the asymmetry in the masses of the constituent quarks can provide further insights on heavy-quark bound-state dynamics.
Extensive predictions of the $\bcq$ excitation spectrum
exist in the literature~\cite{
ChengYQ,
FulcherLP,
Li:2019tbn,
Wang:2022cxy,
Gregory:2009hq,
DowdallRJ,
Mathur:2018epb,
JainP,
EichtenEJ,
Zeng:1994vj,
Gershtein:1994dxw,
Fulcher:1998ka,
Monteiro:2016ijw,
BaldicchiM,
EbertD03,
GodfreyS,
IkhdairSV,
AbdEl-Hady:2005wtn,
Wei:2010zza,
EbertD,
Wang:2012kw,
MonteiroAP2,
Bhat:2017wcl,
SoniNR,
Eichten:2019gig,
Chen:2020ecu,
Martin-Gonzalez:2022qwd,
Li:2022bre,
Li:2023wgq,
Hao:2024nqb},
while the experimental knowledge is relatively limited,
mainly due to the suppressed production rate of the states,
which requires simultaneous production of two heavy quark--antiquark pairs.

The first observation of an excited \bcq state
was reported by the ATLAS experiment~\cite{BPHY-2012-04}
in \pp collision data
collected during Large Hadron Collider (LHC) \RunOne.
The mass of the structure observed in the $\Bcp\pip\pim$ mass spectrum is consistent
with expectations for the first radial excitation of $S$-wave \bcq states ($2S$).
Later, the CMS~\cite{CMS-BPH-18-007,CMS-BPH-19-001} and LHCb~\cite{Aaij:2019ldo} collaborations reported an observation of two distinct signals
compatible with the predictions for the pseudoscalar and vector components of that excitation,
\BcTsp and \BcStTsp, using the full
dataset collected during LHC \RunTwo.
Recently, the LHCb collaboration reported a new observation of a broad peaking structure
in the $\Bcp\gamma$ mass spectrum,
consistent with theory predictions for the lowest excited
$P$-wave \bcq states, \BcOpp~\cite{LHCb:2025uce,LHCb:2025ubr}.
In the case of all of the aforementioned excitations, the mass difference relative to the pseudoscalar ground state, \Bcp,
is between 400 and 600\,\MeV.

Lattice QCD~\cite{Gregory:2009hq,DowdallRJ,Mathur:2018epb} and most of other calculations based on QCD potential models~\cite{
Barik:1993aw,
JainP,
EichtenEJ,
Zeng:1994vj,
Gershtein:1994dxw,
Fulcher:1998ka,
Monteiro:2016ijw,
BaldicchiM,
EbertD03,
GodfreyS,
IkhdairSV,
AbdEl-Hady:2005wtn,
Wei:2010zza,
EbertD,
Wang:2012kw,
MonteiroAP2,
Bhat:2017wcl,
SoniNR,
Eichten:2019gig,
Chen:2020ecu,
Martin-Gonzalez:2022qwd,
Li:2022bre,
Li:2023wgq,
Hao:2024nqb}
predict the mass difference between the lightest excited state of the \bcq system, the vector $\BcStar$ meson,
and the ground state to be between 50 and 70\,\MeV.
The \BcStar meson is expected to decay into \Bcp
by emitting a photon with almost 100\% probability.
No other \bcq excitation is expected to exhibit such a low mass splitting relative to \Bcp.
The small mass difference complicates the experimental search
for the $\BcStar$ due to the very soft energy spectrum of the emitted photon,
which is difficult to detect at LHC experiments.

This Letter presents
the first observation of
a state consistent with theoretical expectations for the $\BcStar$ meson,
decaying into $\Bcp\gamma$
(for simplicity, charge conjugated states are implied from here onwards).
The study is performed with the ATLAS detector~\cite{PERF-2007-01}.
\Bcp mesons are reconstructed in the $\Bcp\to\Jpsi(\to\mumu)\mup\nu_\mu X$ decay chain,
where $X$ corresponds to any other particles which are not reconstructed.
This choice is motivated by the approximately 20 times larger branching fraction compared to the $\Bcp\to\Jpsi\pip$ decay,
used in the earlier \bcq excited state searches at the LHC\@.
The resulting gain in acceptance overcompensates for the complications
related to the presence of the neutrino
and hence the partially reconstructed final state.
Because of the significant \Bcp lifetime the decay candidates are reconstructed from a displaced vertex.
Photons are reconstructed via conversions to \ee in the detector material as vertices of two charged particle tracks.
Typical values of photon transverse momentum relevant to this study are below 1--2\,\GeV
and not accessible to reconstruction approaches based on calorimeter energy clusters.

The ATLAS experiment~\cite{PERF-2007-01} at the LHC is a multipurpose particle detector
with a forward--backward symmetric cylindrical geometry and a near \(4\pi\) coverage in
solid angle.\footnote{\AtlasCoordFootnoteMod}
It consists of an inner tracking detector (ID) surrounded by a thin superconducting solenoid
providing a \qty{2}{\tesla} axial magnetic field, electromagnetic and hadronic calorimeters, and a muon spectrometer (MS).
The ID covers the pseudorapidity range \(|\eta| < 2.5\).
It consists of silicon pixel, silicon microstrip, and transition radiation tracking detectors.
The MS surrounds the calorimeters and is based on
three large superconducting air-core toroidal magnets with eight coils each.
The field integral of the toroids ranges between \num{2.0} and \qty{6.0}{\tesla\metre}
across most of the detector.
The MS includes a system of precision tracking chambers up to \(|\eta| = 2.7\) and fast detectors for triggering up to \(|\eta| = 2.4\).
A two-level trigger system was used to select events~\cite{TRIG-2016-01}.
A software suite~\cite{SOFT-2022-02} is used in data simulation, in the reconstruction
and analysis of real and simulated data, in detector operations, and in the trigger and data acquisition
systems of the experiment.

This analysis is based on the full \RunTwo dataset of $\sqs = 13$\,\TeV \pp collisions collected by ATLAS between 2015 and 2018
at the LHC.
The data are required to satisfy criteria ensuring that the
detector was in good operating condition~\cite{DAPR-2018-01}.
Events are triggered on the basis of the muons present in the final state of the targeted decay channel~\cite{TRIG-2018-01}.
Several trigger configurations were
used to maximise the signal yield.
Most of the triggers were based on the identification of two muons,
with various muon transverse momentum thresholds from 4 to 11\,\GeV,
requiring opposite charge for the muons in the pair, to form a good vertex and to have an invariant mass compatible with
a $\Jpsi\to\mumu$ decay.
Other triggers required the presence of a third muon,
typically lowering the thresholds down to 4\,\GeV for all three while still keeping acceptable rates.
The total trigger efficiency is above 60\% for \Bcp mesons with \pT above 25\,\GeV.

To model inelastic \pp events containing the $\BcStar\to\Bcp\gamma$ decays,
followed by the decays $\Bcp\to\Jpsi\mup\nu_\mu X$,
large samples of Monte Carlo (MC) simulated events
were prepared using BCVEGPY~2.2 generator~\cite{Chang:2015qea} to model \bcq state production,
interfaced to \Pythia[8.244]~\cite{Sjostrand:2014zea} to simulate
parton showers and hadronisation, and \EVTGEN~\cite{Lange:2001uf} to model heavy-flavour decays.
All generated MC events undergo the full ATLAS detector simulation~\cite{SOFT-2022-02}
using \GEANT~\cite{Agostinelli:2002hh}.

Events are required to contain at least three muons reconstructed using
information from the MS and the ID and satisfying the \emph{Loose}
identification working point~\cite{MUON-2018-03}.
The $\Jpsi$ candidates are built from two oppositely charged muons,
which are required to have $\pT(\mu^\pm)>4$\,\GeV and $|\eta(\mu^\pm)|<2.4$,
and that have their tracks fitted to a common vertex~\cite{Kostyukhin:2003daa}.
Candidates in the dimuon invariant mass window $2.8\,\GeV < m(\mumu) < 3.4\,\GeV$ are retained
for further analysis.
The third muon track with $\pT(\mu^\pm)>3$\,\GeV and $|\eta(\mu^\pm)|<2.5$
is fitted to a common vertex with the muon tracks of the \Jpsi candidate.
In this fit the $\Jpsi$ candidate mass is constrained to the world average value~\cite{PDG2024}.
The combination of the \Jpsi candidate with the third muon is referred
to as the \Bcp candidate below.
Because of the presence of the neutrino in the \Bcp decay,
the invariant mass of the signal candidates is expected to have a smooth
distribution in the kinematically allowed range of roughly 3.2--6.3\,\GeV.
Primary vertices (PV) are reconstructed from at least two charged-particle tracks
with $\pT > 0.5$\,\GeV and $|\eta|<2.5$~\cite{PERF-2015-01,ATL-PHYS-PUB-2015-026}.
The \Bcp candidate is associated to a PV by choosing
the one giving the smallest three-dimensional impact parameter of the $\Bcp$ candidate trajectory.
The position of the PV is determined from a refit of the vertex, following the removal of the muon
tracks used to reconstruct the $\Bcp$ candidate.
The \Bcp vertex fit quality must be $\chi^2(\Bcp)/\ndf<2$,
where the number of degrees of freedom $\ndf=4$.
The contribution of the third muon track to the $\chi^2(\Bcp)$ is required to be less than 4,
to suppress background from a genuine \Jpsi combined with a randomly-associated muon.
The projection of the $\Bcp$ candidate transverse momentum
on the direction from the PV to the candidate decay vertex
is required to be positive, to further suppress the background from prompt \Jpsi, produced at the PV.
The \Bcp candidates are required to satisfy the criterion
$\pT(\Bcp)/\sum{\pT(\text{track})}>0.2$,
where the scalar sum in the denominator is taken over all tracks with $\pT > 0.5$\,\GeV
originating from the PV, including the tracks of the \Bcp candidate.
The \Bcp candidates must have $\pT(\Bcp)>25$\,\GeV and $|\eta(\Bcp)|<2.4$.
If the same triplet of muons allows the formation of two \Bcp candidates passing the selection above
(with different muons considered as the \Jpsi daughters), only the candidate with the lowest
$\chi^2(\Bcp)$ is retained.
Multiple \Bcp candidates in a single event are kept if they share up to two muons.
This happens in less than 0.2\% of events, for both data and simulated event samples.

\begin{figure}[htbp]
\centering
\includegraphics[width=0.49\textwidth]{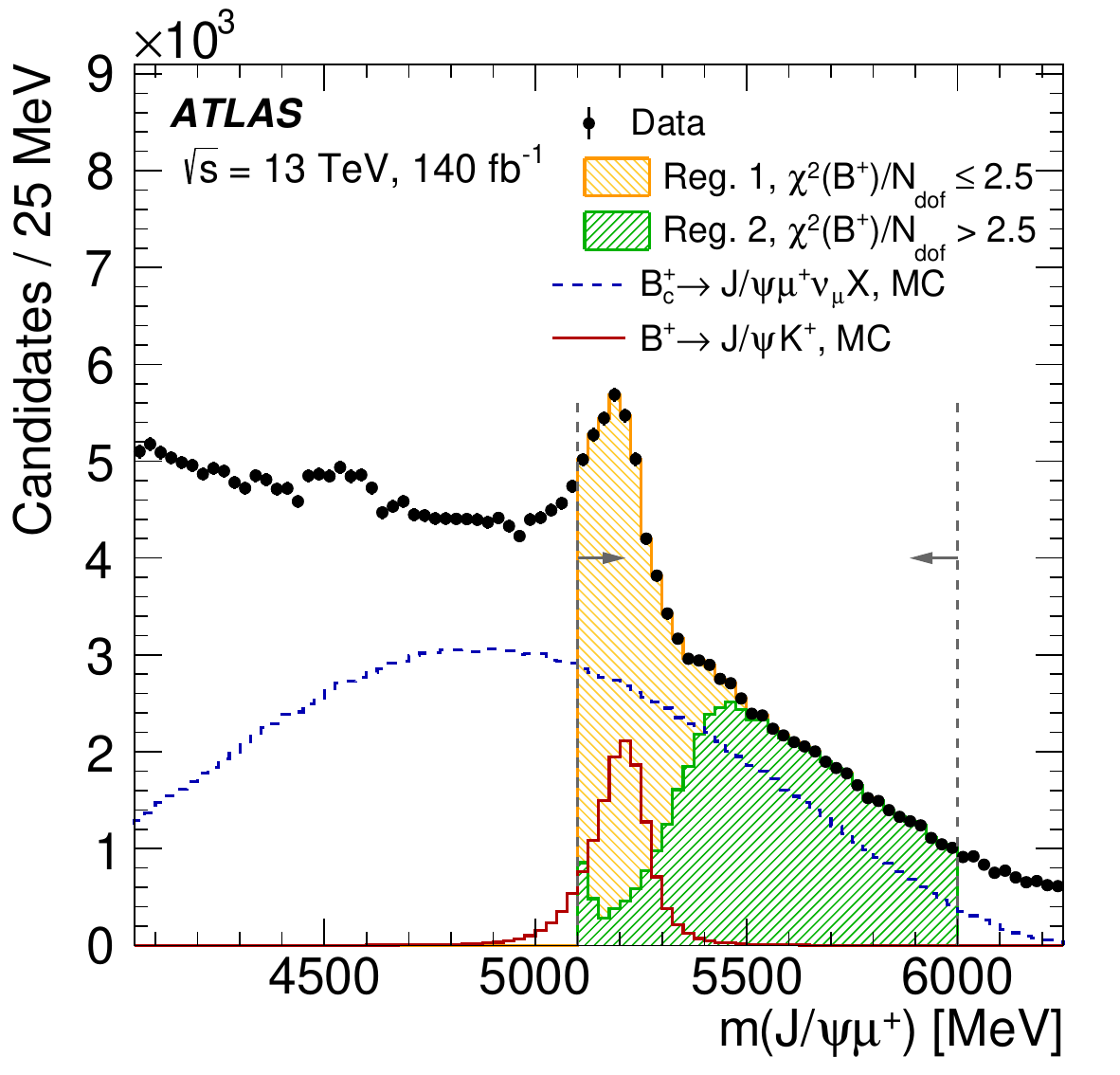}
\caption{
The distribution of \Bcp candidate invariant mass.
The colour-filled areas show the $\Bcp$ candidates in the mass window
selected for further analysis:
from (orange) \dsone\ -- candidates with $\chi^2(B^+)/\ndf\leq 2.5$,
and (green) \dstwo\ -- candidates with $\chi^2(\Bp)/\ndf>2.5$,
where the quantity $\chi^2(\Bp)$ is defined in the text.
The blue dashed line shows the expected contribution of $\Bcp\to\Jpsi\mup\nu_\mu X$ decays,
with the normalisation calculated from the fitted \BcStar signal yield and simulated ratio of
reconstruction efficiencies of \BcStar and \Bcp candidates,
and extrapolated
to the total expected \Bcp inclusive production.
The red solid line shows the contribution of $\Bp\to\Jpsi\Kp$ decay.
The grey vertical dashed lines and arrows indicate the mass range of \Bcp candidates used for \BcStar reconstruction.
\label{fig:massct}
}
\end{figure}

Figure~\ref{fig:massct} shows
the distribution of invariant mass, $m(\Jpsi\mup)$,
for all $\Bcp$ candidates with
$4.0 < m(\Jpsi\mup) < 6.25$\,\GeV.
Studies performed with simulation suggest that
the distribution
contains sizeable contributions
from signal $\Bcp\to\Jpsi\mup\nu_\mu X$ decays
and
from background $\Jpsi$ mesons combined with either
a true muon or a hadron mis-identified as a muon,
due to punch-through to the MS or decays in flight.
A narrow peak with $m(\Jpsi\mup)\sim 5.2$\,\GeV is produced
by $\Bp\to\Jpsi\Kp$ decays where $\Kp$ is mis-identified
as a muon.
The structure near 4.5\,\GeV arises due to various $\Bp\to(\psi/\chi)\Kp\to\Jpsi\Kp X$ decays
with the same kaon mis-identification and
where some of the decay products of the heavier charmonium state, such as $\psip$ or $\chi_{c\mathrm{J}}$, are not reconstructed.
Many other partially reconstructed $B\to\Jpsi X$ decays can contribute
to the mass range below about 5\,\GeV.
Figure~\ref{fig:massct} also shows the expected contribution from the $\Bcp\to\Jpsi\mup\nu_\mu X$ signal.
Its shape is taken from the simulation
and its normalisation is obtained from the fitted yield of \BcStar signal in data discussed below.
The normalisation also uses the ratio of \BcStar and \Bcp reconstruction efficiencies, obtained from simulation,
and the assumption that 3/4 of the inclusive \Bcp mesons are produced in decays of the vector \BcStar state.
The latter assumption is motivated by the ratio of the spin states involved in the production (naive spin counting).
It is used only for illustrative purposes and no further aspects of the analysis rely on this normalisation.
The contribution of the $\Bp\to\Jpsi\Kp$ decay with the shape taken from the simulation and the normalisation
extracted by fitting the invariant mass distribution assuming a kaon
mass hypothesis for the third muon track amounts to about 15000 candidates and is also shown in Figure~\ref{fig:massct}.

To increase the fraction of $\Bcp\to\Jpsi\mup\nu_\mu X$ decays and suppress contributions
from other partially reconstructed $b$-hadron decays,
only $\bcp$ candidates with
$5.1 < m(\Jpsi\mup) < 6.0$\,\GeV are kept for further
analysis.
The contribution from $\Bp\to\Jpsi\Kp$ decays
can be isolated by performing a modified 3-muon vertex fit as follows.
A kaon mass hypothesis is assigned to the third muon track, and the triplet
is refitted to a common vertex with the following constraints: the \Jpsi candidate muon pair and whole triplet
masses are constrained to the world average values for \Jpsi and $B^+$ masses~\cite{PDG2024}, respectively,
and the reconstructed candidate
momentum is required to point back to the PV\@.
The quality of this modified vertex fit, $\chi^2(\Bp)/\ndf$
with $\ndf=7$,
is used to define two non-overlapping regions:
\dsone includes candidates with $\chi^2(B^+)/\ndf\leq 2.5$ and is expected to contain
a significant contribution from $\Bp\to\Jpsi\Kp$
and from combinatorial background;
\dstwo includes candidates with $\chi^2(B^+)/\ndf>2.5$.
According to the simulation, the expected residual $\Bp\to\Jpsi\Kp$ contribution in \dstwo
does not exceed 2\% of its total contribution to both regions.
Both regions are kept for further analysis and are shown in Figure~\ref{fig:massct} for the data sample.

Candidates for \BcStar mesons decaying into $\Bcp\gamma$ are reconstructed
by combining \Bcp candidates with a soft photon.
To reconstruct the photons, the process of photon conversion in the detector material is exploited.
Pairs of oppositely charged ID tracks (both assigned an electron mass hypothesis)
are used for the reconstruction of conversion vertices.
To increase the acceptance for the low-energy photons,
each track is required to have $\pT(e^\pm)>100$\,\MeV and $|\eta(e^\pm)|<2.5$.
This \pT threshold is much lower than a typical value of 400--500\,\MeV
used in most ATLAS data analyses~\cite{PERF-2015-08,IDTR-2021-01}.
A dedicated track reconstruction setup similar to that used in measurements with minimum-bias data~\cite{STDM-2015-17}
was employed, with certain computational adaptions
made for the track multiplicity environment of nominal \RunTwo data-taking conditions.
This dedicated reconstruction was run on a preselected set of events~\cite{Barberis:2022tsf} containing a \Bcp candidate.
Track pairs are fitted to a common vertex,
with the quality of the fit with one degree of freedom required to satisfy
$\chi^2 < 3$.
The vertex radial and longitudinal positions are required to be $20 < \Rxy(\ee) < 125$\,mm and $|z(\ee)|< 425$\,mm, respectively.
This geometrical region roughly corresponds to the volume occupied by the beampipe and barrel layers of the silicon pixel detector.
Photon conversion candidates are required to have an invariant mass $\mee < 50$\,\MeV
and transverse momentum
with respect to the direction from the PV to the conversion vertex $\ptrel < 20$\,\MeV.
To further suppress fake conversion candidates,
a combined selection criterion
is applied in the two-dimensional space of \mee and \ptrel,
where the signal is concentrated towards small values of both variables,
and the background dominates at higher values.
The reconstruction and selection efficiency for photons with \pT above 200\,\MeV is
approximately 0.4\% in the \BcStar signal simulation.
This is defined mainly by the radiation thickness of the traversed material,
acceptance of the $\pT(e^\pm)$ selection and the efficiency of low-energy electron track reconstruction.

Finally, the invariant mass of the muon triplet and the photon candidate with a kaon mass hypothesis assigned to the third muon
is required to be $m(\Jpsi\Kp\ee) < 6050$\,\MeV.
This requirement suppresses possible contributions from fully reconstructed $\Bcp\to\Jpsi\Kp$ and $\Bcp\to\Jpsi\pip$
decays with soft photons from final-state radiation.
If more than one \BcStar candidate passes the selection in a single event, all of them are kept.
This happens in about 2\% of events in both data and \BcStar signal simulation.

\begin{figure}[htbp]
\centering
\subfloat[][]{
\includegraphics[width=0.49\textwidth]{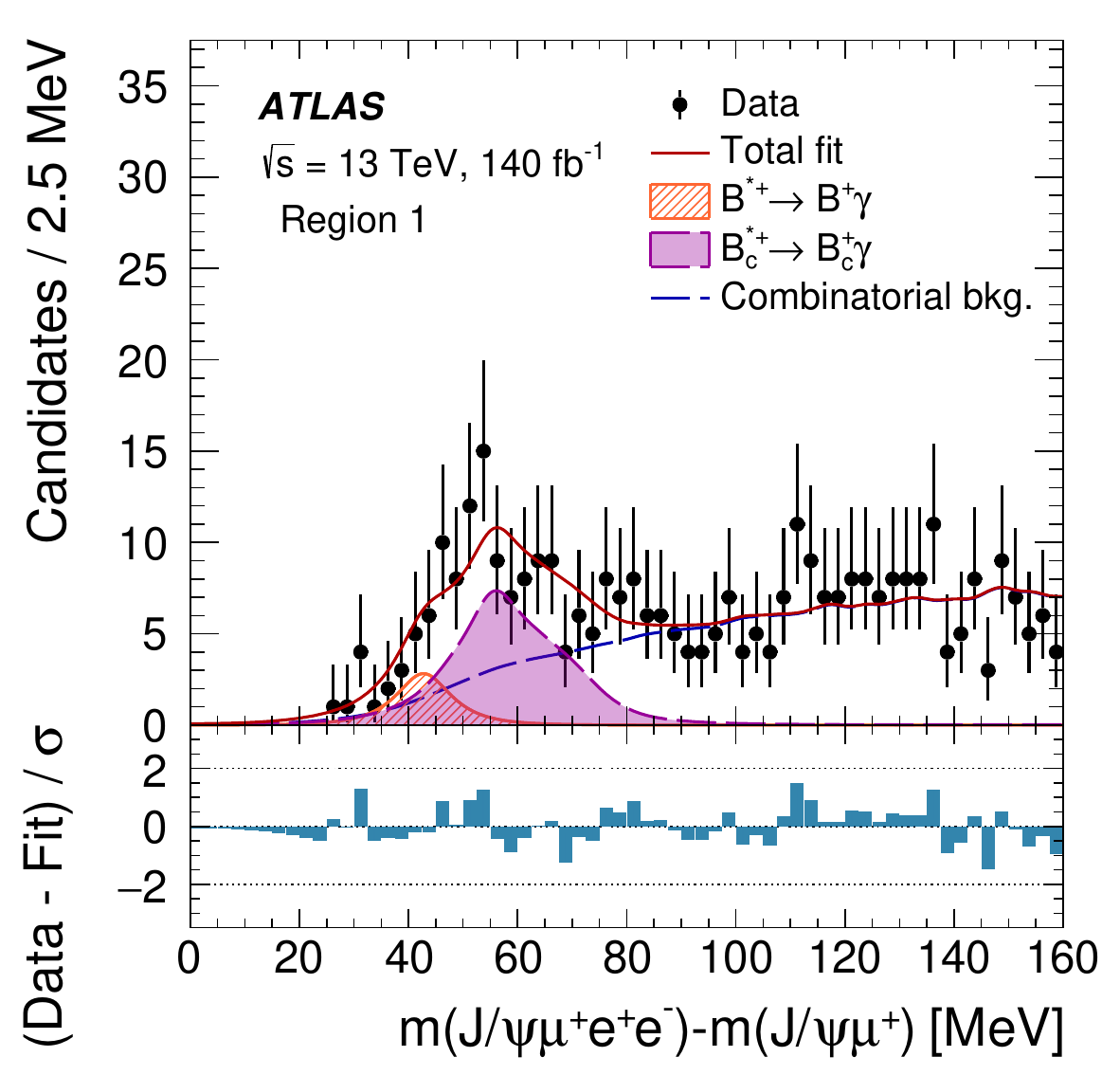}
\label{fig:fitBcStar:q_DS1}
}
\subfloat[][]{
\includegraphics[width=0.49\textwidth]{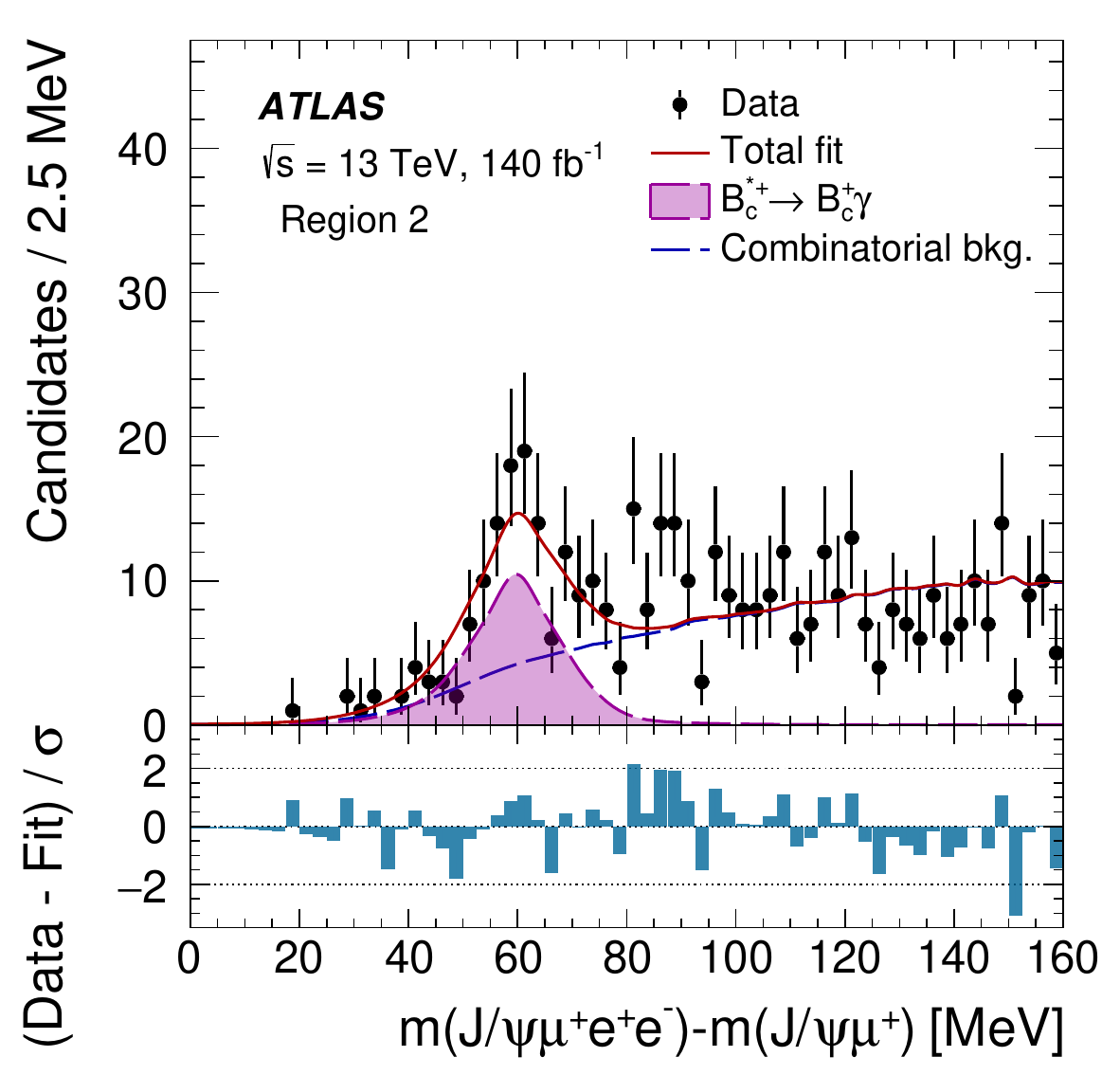}
\label{fig:fitBcStar:q_DS2}
}
\\
\caption{
Distributions of the mass difference for the selected \BcStar candidates in \protect\subref{fig:fitBcStar:q_DS1}~\dsone
and \protect\subref{fig:fitBcStar:q_DS2}~\dstwo.
Data are shown as points, and the overall result of the fit is given by the red solid line.
The purple filled area shows the \BcStar signal contribution,
and the orange hatched area corresponds to the \BpStar background.
The combinatorial background is shown with the blue dashed line.
The bottom panels show the pulls, defined as the difference between the data and the fit function
divided by the statistical uncertainty of the data point.
\label{fig:fitBcStar}
}
\end{figure}

The signal of the $\BcStar\to\Bcp\gamma$ decay is identified using the distribution of the mass difference
$\dm = m(\Jpsi\mup\ee)-m(\Jpsi\mup)$.
Figure~\ref{fig:fitBcStar} shows the \dm distribution for the selected \BcStar candidates
in \dsone and \dstwo.
This distribution features a structure in the region $40\,\MeV <\dm<70$\,\MeV
in both regions.
The observed excess of events above the expected, smoothly rising, background distribution
in \dstwo is attributed exclusively to the $\BcStar\to\Bcp\gamma$ decay signal,
while in \dsone the same signal can be accompanied by $\Bpst\to\Bp\gamma$ decays,
which are expected to produce a peaking contribution
with a \dm value close to $\dmbps = \mbps - \mbp = 45.34\pm 0.20$\,\MeV~\cite{PDG2024}.
The expected $\Bpst$ contamination in \dstwo is negligible.

An extended unbinned maximum-likelihood fit to the $\dm$ distribution from 0 to 160\,\MeV
(full \dm range)
is performed simultaneously to the two regions.
The $\BcStar\to\Bcp\gamma$ decay signal contribution is
modelled with a template made from the simulated event samples
with the adaptive Gaussian kernel density estimation (KDE) technique~\cite{Cranmer:2000du}.
Separate templates are used for \dsone and 2.
The dependence of the template
on the true mass difference value, $\dmbcs = \mbcs-\mbc$,
was studied using simulated signal samples with different values of \mbcs.
Due to the lost neutrino in the \Bcp decay final state
the mean \dm position of the template is shifted by a few \MeV relative to \dmbcs
and the templates are broader than would be expected from the photon energy resolution alone.
In the fit to data, the \dm values of the simulated events used to build the template are rescaled to
$\dm\times\dmbcs/\dmbcs^{\text{nominal}}$,
where \dmbcs is a free parameter of the fit
and $\dmbcs^{\text{nominal}}$ is the true mass difference value used in the simulated sample.
This linear scaling was shown to well reproduce the signal \dm shape for the \dmbcs values in the 60--70\,\MeV range.

The $\BpStar\to\Bp\gamma$ decay is modelled
with a modified Gaussian function,
$G_\textrm{mod}\propto \exp\left(-0.5\times x^{2-x/(x+2)}\right)$~\cite{Chekanov:2005cf,BPHY-2011-02},
where $x=|\Delta m -\Delta m_{\BpStar}|/\sigma_{\BpStar}$ with the mean $\dmbps$ and width
$\sigma_{\BpStar}$ fixed
to the values obtained from simulated samples.
This component only contributes to \dsone and is absent in \dstwo.

The background is dominated by the combinatorial component and includes a small contribution
of photons produced from final-state radiation in $\Bcp\to\Jpsi\mup\nu_\mu X$ decays.
According to the simulation,
the latter contribution does not exceed 5\% of the total background
and its \dm shape is consistent with that of the combinatorial background,
therefore it is not considered separately.
For the background modelling, a data-driven method, referred to as event mixing below, is applied.
\BcStar candidates in data with $80 < \dm < 160$\,\MeV (partial \dm range) are used to build synthetic background candidates
by combining the \Bcp and photon from different events.
The \dm range below 80\,\MeV is excluded to avoid contamination from the signal.
The use of the partial \dm range biases the background $\pT(\gamma)$ spectrum,
as does the mixing procedure itself, favouring ``mixed'' candidates with low
$\pT(\gamma)$ which are more likely to have $\dm < 160$\,\MeV.
These biases should be corrected since the $\pT(\gamma)$ spectrum is highly correlated with the \dm shape.
Corrections for the partial \dm range are derived
using the \BcStar candidates in the signal simulation where a reconstructed photon
is not matched to the true photon from \BcStar (unmatched candidates),
by taking the ratio of the $\pT(\gamma)$ spectra of the candidates in the partial and full \dm ranges.
The bias due to the mixing procedure itself is corrected by the ratio of the $\pT(\gamma)$
spectra in data candidates used for the mixing and in the ``mixed'' candidates in the partial \dm range.
The \dm templates for the combinatorial background are made from the corrected ``mixed'' candidates
using the KDE technique.
Separate templates are made for the \dsone and \dstwo by
using the corresponding \Bcp candidate selections.
The background yields in each region are allowed to float independently in the fit.

The total \BcStar signal yield in \dsone and 2, $N_{\BcStar}$,
and \BpStar yield in \dsone, $N_{\BpStar}$, are free parameters of the fit.
The ratio of the \BcStar yields in \dsone and 2, \NDSratio,
is fixed to the value extracted from the signal simulation.

The results of the simultaneous fit
are shown in Figure~\ref{fig:fitBcStar}.
The total \BcStar signal yield
in data is found to be $N_{\BcStar} = 163 \pm 19$, where this uncertainty is statistical only.
The yield of the \BpStar, $N_{\BpStar} = 17 \pm 8$, is consistent with expectations based on the
amount of \Bp mesons in the peak in the $m(\Jpsi\mup)$ distribution
and the simulated efficiency for $\Bpst\to\Bp\gamma$ events.
The signal significance relative to the background-only hypothesis is evaluated
using the asymptotic distribution of the profile likelihood ratio~\cite{Cowan:2010js}.
The method of Ref.~\cite{Gross:2010qma} is used to account for the look-elsewhere effect.
Both the local and global significance exceed 10 standard deviations assuming the nominal fit model
and stay above 8 standard deviations for all systematic variations described below.
The fit yields a \BcStar mass difference value of $\dmbcs = 64.5 \pm 1.4$\,\MeV,
where the uncertainty is statistical.
Separate fits of \dsone and 2 individually yield consistent \dmbcs values.

Systematic uncertainties in the value of \dmbcs arise due to detector-related and modelling effects,
affecting the shape of signal templates used in the \dm fit,
and the signal extraction procedure itself.
Each source of systematic uncertainty is considered individually,
either by repeating the analysis
with an appropriate change in the methodology implemented,
or by means of MC pseudo-experiments to assess the sensitivity of the fit to the input assumptions.
The uncertainty sources that have a significant impact on the \dmbcs measurement
are listed below and their contributions are summarised in Table~\ref{tab:syst}.
Contributions of individual sources are treated as uncorrelated and added in quadrature.

The largest systematic uncertainty is associated with the ID
material modelling and photon momentum scale and resolution in simulation.
Limited knowledge of the actual material distribution in the ID volume can affect
the modelling the photon conversion probability and the efficiency of photon conversion reconstruction.
This is investigated using
a set of alternative \BcStar signal simulations
with varying detector geometry descriptions or \GEANT physics models~\cite{PERF-2015-07}.
The accuracy of the simulation's modelling of the photon momentum scale and resolution is evaluated
in a dedicated study with the $\BpStar\to\Bp\gamma$ process with subsequent $\Bp\to\Jpsi\Kp$
decay used as a control channel.
The difference in the reconstructed position and width of the \BpStar peak in the distribution of
$\dm^\prime = m(\Jpsi\Kp\ee)-m(\Jpsi\Kp)$ between data and simulation
indicates a small discrepancy in both the simulated momentum scale and resolution.
An associated systematic uncertainty is estimated by
adjusting the detector \dm resolution
and the photon \pT of the simulated candidates
by factors that impose consistency of the \BpStar peak width and position in the simulation and data.
The effect of mass resolution ($-0.85$\,\MeV) and the momentum scale ($-0.74$\,\MeV) are added in quadrature
resulting in a \dmbcs shift of $-1.13$\,\MeV.
Since they are partially caused by the uncertainty of the ID material modelling,
the latter is not added in quadrature, but considered only for the positive-side \dmbcs uncertainty.

To evaluate the effect of the $\Bcp$ production modelling,
the simulated $\pT$ and $|\eta|$ spectra of the \Bcp in the signal simulation are varied
based on the comparison between data and simulation performed with $\Bcp\to\Jpsi\pip$ channel~\cite{BPHY-2018-08}.
The effect of the modelling of the $\Bcp\to\Jpsi\mup\nu_\mu$ decay,
which uses the decay form factors outlined in Refs.~\cite{Kiselev:1993ea,Kiselev:1999sc,Kiselev:2002vz}
in the nominal simulation,
is evaluated by using
alternative models based on Refs.~\cite{Ebert:2003cn}, \cite{Scora:1995ty}, or the phase-space model.
In addition to the dominant $\Bcp\to\Jpsi\mup\nu_\mu$ channel, other decay chains yielding the \Jpsi and a muon in the \Bcp
decay final state contribute to the inclusive $\Bcp\to\Jpsi\mup\nu_\mu X$ signal and are included in the nominal simulation.
The total fraction of those decays in the simulated sample (about 6\% after the selection) is varied between zero and
twice the nominal value.
These variations imply an effect on the \NDSratio ratio, fixed in the data fit.
Another variation of the fit, where this ratio is treated as a free parameter,
is used to conservatively account for a possible residual uncertainty associated with its modelling.
Such a fit yields a value consistent with the expectation from the simulation.
Muon reconstruction~\cite{MUON-2018-03} and trigger efficiency~\cite{TRIG-2018-01} uncertainties
are propagated to the signal \dm template shapes.

Several alternative approaches to describe the combinatorial background are used
to estimate the systematic uncertainties associated with the modelling of its shape.
Varying the event mixing procedure by keeping the $z$ projection of the photon momentum from the original \BcStar
candidate, or by using exclusively the unmatched \Bcst candidates in the signal MC sample
to correct the $\pT(\gamma)$ spectrum biases, yields tiny variations of the fit results.
Using an analytical function fit to the \dm shape of the unmatched candidates for the
background modelling instead of the event mixing
results in a sizeable change in the background shape,
making this second largest contribution to the overall systematic uncertainty.
Across all systematic variations, the fit with this alternative background description yields the smallest estimate
of the signal significance and is associated with a signal yield reduction of about 15\%.

\begin{table}[h]
\centering
\caption{
Systematic uncertainties associated with the measured value of \dmbcs.
}
\renewcommand{\arraystretch}{1.4}  %
\begin{tabular}{l|c}
\toprule
Source & Uncertainty [\MeV{}] \\
\midrule
Material modelling and photon momentum scale \& resolution &  $^{+0.44}_{-1.13}$    \\
\Bcp production modelling                                  &  \footnotesize{$\pm 0.19$}   \\
$\Bcp\to\Jpsi\mup\nu_\mu X$ decay modelling                       &  \footnotesize{$\pm 0.05$}   \\
\NDSratio ratio                                            &  \footnotesize{$\pm 0.12$} \\
Muon reconstruction and trigger                            &  \footnotesize{$\pm 0.16$} \\
Background shape modelling                                 &  \footnotesize{$\pm 0.79$} \\
\midrule
Total                                                      &  $^{+0.95}_{-1.41}$   \\
\bottomrule
\end{tabular}
\label{tab:syst}
\end{table}

In summary, a new state decaying into \Bcp and a photon was observed by the ATLAS experiment
at the LHC using \pp collision
data corresponding to an integrated luminosity of $140$\,\ifb\
at 13\,\TeV centre-of-mass energy.
The study was performed using the $\Bcp\to\Jpsi\mup\nu_\mu X$ decay channel
and an innovative approach to build photon conversion vertices from low-\pT tracks
reconstructed in a preselected event sample.
The mass difference between the new state and ground-state $\Bcp$ meson
is measured to be $\dmbcs = 64.5\pm 1.4\stat^{+1.0}_{-1.4}\syst$\,\MeV.
This value matches the theory expectations
for the lowest vector state of the \bcq system, the \BcStar meson,
which are generally in the range of 50--70\,\MeV~\cite{Gregory:2009hq,DowdallRJ,Mathur:2018epb,
Barik:1993aw,
JainP,
EichtenEJ,
Zeng:1994vj,
Gershtein:1994dxw,
Fulcher:1998ka,
Monteiro:2016ijw,
BaldicchiM,
EbertD03,
GodfreyS,
IkhdairSV,
AbdEl-Hady:2005wtn,
Wei:2010zza,
EbertD,
Wang:2012kw,
MonteiroAP2,
Bhat:2017wcl,
SoniNR,
Eichten:2019gig,
Chen:2020ecu,
Martin-Gonzalez:2022qwd,
Li:2022bre,
Li:2023wgq,
Hao:2024nqb}.
However, it is larger than the values from the most recent precise
lattice QCD calculations~\cite{Gregory:2009hq,DowdallRJ,Mathur:2018epb}
which are below 60\,\MeV.
The measured \dmbcs value corresponds to a mass for the new state of
$m(\BcStar)=6339.0\pm 1.4\stat^{+1.0}_{-1.4}\syst\pm 0.3\errmbc$\,\MeV,
where the third uncertainty is due to the precision of the $B_c^+$ meson mass knowledge~\cite{PDG2024}.
The new state is observed with a significance that exceeds 8 standard deviations.
\section*{Acknowledgements}
%

%

%
%

%
%

We thank CERN for the very successful operation of the LHC and its injectors, as well as the support staff at
CERN and at our institutions worldwide without whom ATLAS could not be operated efficiently.

The crucial computing support from all WLCG partners is acknowledged gratefully, in particular from CERN, the ATLAS Tier-1 facilities at TRIUMF/SFU (Canada), NDGF (Denmark, Norway, Sweden), CC-IN2P3 (France), KIT/GridKA (Germany), INFN-CNAF (Italy), NL-T1 (Netherlands), PIC (Spain), RAL (UK) and BNL (USA), the Tier-2 facilities worldwide and large non-WLCG resource providers. Major contributors of computing resources are listed in Ref.~\cite{ATL-SOFT-PUB-2026-001}.

We gratefully acknowledge the support of ANPCyT, Argentina; YerPhI, Armenia; ARC, Australia; BMWFW and FWF, Austria; ANAS, Azerbaijan; CNPq and FAPESP, Brazil; NSERC, NRC and CFI, Canada; CERN; ANID, Chile; CAS, MOST, NSFC and FRFCU, China; Minciencias, Colombia; MEYS CR, Czech Republic; DNRF and DNSRC, Denmark; IN2P3-CNRS and CEA-DRF/IRFU, France; SRNSFG, Georgia; BMFTR, HGF and MPG, Germany; GSRI, Greece; RGC and Hong Kong SAR, China; ICHEP and Academy of Sciences and Humanities, Israel; INFN, Italy; MEXT and JSPS, Japan; CNRST, Morocco; NWO, Netherlands; RCN, Norway; MNiSW, Poland; FCT, Portugal; MNE/IFA, Romania; MSTDI, Serbia; MSSR, Slovakia; ARIS and MVZI, Slovenia; DSI/NRF, South Africa; MICIU/AEI, Spain; SRC and Wallenberg Foundation, Sweden; SERI, SNSF and Cantons of Bern and Geneva, Switzerland; NSTC, Taipei; TENMAK, T\"urkiye; STFC/UKRI, United Kingdom; DOE and NSF, United States of America.

Individual groups and members have received support from BCKDF, CANARIE, CRC and DRAC, Canada; CERN-CZ, FORTE and PRIMUS, Czech Republic; COST, ERC, ERDF, Horizon 2020 and Marie Sk{\l}odowska-Curie Actions, European Union; Investissements d'Avenir Labex, Investissements d'Avenir Idex and ANR, France; DFG and AvH Foundation, Germany; Herakleitos, Thales and Aristeia programmes co-financed by EU-ESF and the Greek NSRF, Greece; BSF-NSF and MINERVA, Israel; NCN and NAWA, Poland; La Caixa Banking Foundation, CERCA and AGAUR programs from Generalitat de Catalunya and PROMETEO and GenT Programmes Generalitat Valenciana, Spain; G\"{o}ran Gustafssons Stiftelse, Sweden; The Royal Society and Leverhulme Trust, United Kingdom; Eric and Wendy Schmidt Fund for Strategic Innovation, United States of America.

In addition, individual members wish to acknowledge support from Chile: Agencia Nacional de Investigaci\'on y Desarrollo (ANID FONDECYT reg. 1230987, FONDECYT 1230812, FONDECYT 1240864, Fondecyt 3240661, Fondecyt Regular 1240721); China: Fundamental Research Funds for the Central Universities (010-63263105), Chinese Ministry of Science and Technology (MOST-2023YFA1605700, MOST-2023YFA1609300), National Natural Science Foundation of China (NSFC 12275265, NSFC-W2543005); Czech Republic: Czech Science Foundation (GACR - 24-11373S), Ministry of Education Youth and Sports (ERC-CZ-LL2327, FORTE CZ.02.01.01/00/22\_008/0004632); EU: H2020 European Research Council (ERC - 101002463); European Union: European Research Council (BARD No. 101116429, ERC - 101219398, ERC 101089007), European Regional Development Fund (HE COFUND GA No.101081355, ERDF), Marie Sklodowska-Curie Actions (GAP-101168829); France: Agence Nationale de la Recherche (ANR-21-CE31-0013, ANR-22-EDIR-0002, ANR-24-CE31-0504-01); Germany: Deutsche Forschungsgemeinschaft (DFG - 469666862); China: Research Grants Council (GRF); Italy: Istituto Nazionale di Fisica Nucleare (LHC-MIUR - 28003/2025), Ministero dell'Università e della Ricerca (NextGenEU I53D23000820006 M4C2.1.1, SOE2024\_0000023); Japan: Japan Society for the Promotion of Science (JSPS KAKENHI  JP25H0063, JSPS KAKENHI JP22H04944, JSPS KAKENHI JP24K23939, JSPS KAKENHI JP24KK0251, JSPS KAKENHI JP25H00650, JSPS KAKENHI JP25H01291, JSPS KAKENHI JP25K01011, JSPS KAKENHI JP25K01023, JSPS KAKENHI JP25KK0047); Poland: Polish National Science Centre (NCN 2021/42/E/ST2/00350, NCN OPUS 2023/51/B/ST2/02507, NCN OPUS nr 2022/47/B/ST2/03059, NCN UMO-2019/34/E/ST2/00393, UMO-2022/47/O/ST2/00148, UMO-2023/49/B/ST2/04085, UMO-2023/51/B/ST2/00920, UMO-2024/53/N/ST2/00869); Spain: Agència de Gestió d'Ajuts Universitaris i de Recerca. (AGAUR - 2023 BP 00141), Ministry of Science and Innovation (RYC2019-028510-I, RYC2020-030254-I, RYC2021-031273-I, RYC2022-038164-I), Ministerio de Ciencia, Innovación y Universidades/Agencia Estatal de Investigaci\'on (EU NextGenerationEU (PRTR-C17.I1), PID2022-142604OB-C22); Sweden: Carl Trygger Foundation (Carl Trygger Foundation CTS 22:2312), Swedish Research Council (Swedish Research Council 2023-04654, VR 2021-03651, VR 2022-03845, VR 2022-04683, VR 2023-03403, VR 2024-05451, VR 2025-05940), Knut and Alice Wallenberg Foundation (KAW 2023.0366); United Kingdom: The Binks Trust, Royal Society (NIF-R1-231091); United States of America: U.S. Department of Energy (ECA DE-AC02-76SF00515), John Templeton Foundation (John Templeton Foundation 63206), Neubauer Family Foundation.

%
%


%
%
%
%
%
%

%
%

%
%
\printbibliography
\clearpage
 
\begin{flushleft}
\hypersetup{urlcolor=black}
{\Large The ATLAS Collaboration}

\bigskip

\AtlasOrcid[0000-0002-6665-4934]{G.~Aad}$^\textrm{\scriptsize 103}$,
\AtlasOrcid[0000-0001-7616-1554]{E.~Aakvaag}$^\textrm{\scriptsize 17}$,
\AtlasOrcid[0000-0002-5888-2734]{B.~Abbott}$^\textrm{\scriptsize 122}$,
\AtlasOrcid[0000-0002-0287-5869]{S.~Abdelhameed}$^\textrm{\scriptsize 84b}$,
\AtlasOrcid[0000-0002-1002-1652]{K.~Abeling}$^\textrm{\scriptsize 55}$,
\AtlasOrcid[0000-0001-5763-2760]{N.J.~Abicht}$^\textrm{\scriptsize 49}$,
\AtlasOrcid[0000-0002-8496-9294]{S.H.~Abidi}$^\textrm{\scriptsize 30}$,
\AtlasOrcid[0009-0003-6578-220X]{M.~Aboelela}$^\textrm{\scriptsize 45}$,
\AtlasOrcid[0000-0002-9987-2292]{A.~Aboulhorma}$^\textrm{\scriptsize 36e}$,
\AtlasOrcid[0000-0001-5329-6640]{H.~Abramowicz}$^\textrm{\scriptsize 155}$
% ...
\bigskip
\\

% ...
$^{17}$Department for Physics and Technology, University of Bergen, Bergen; Norway.\\
% ...
$^{30}$Physics Department, Brookhaven National Laboratory, Upton NY; United States of America.\\
% ...
$^{36}$$^{(a)}$Facult\'e des Sciences Ain Chock, Universit\'e Hassan II de Casablanca;$^{(b)}$Facult\'{e} des Sciences, Universit\'{e} Ibn-Tofail, K\'{e}nitra;$^{(c)}$Facult\'e des Sciences Semlalia, Universit\'e Cadi Ayyad, LPHEA-Marrakech;$^{(d)}$LPMR, Facult\'e des Sciences, Universit\'e Mohamed Premier, Oujda;$^{(e)}$Facult\'e des sciences, Universit\'e Mohammed V, Rabat;$^{(f)}$Institute of Applied Physics, Mohammed VI Polytechnic University, Ben Guerir; Morocco.\\
% ...
$^{45}$Physics Department, Southern Methodist University, Dallas TX; United States of America.\\
% ...
$^{49}$Fakult\"{a}t Physik, Technische Universit{\"a}t Dortmund, Dortmund; Germany.\\
% ...
$^{55}$II. Physikalisches Institut, Georg-August-Universit\"{a}t G\"ottingen, G\"ottingen; Germany.\\
% ...
$^{84}$$^{(a)}$Khalifa University of Science and Technology, Abu Dhabi;$^{(b)}$New York University Abu Dhabi, Abu Dhabi;$^{(c)}$United Arab Emirates University, Al Ain;$^{(d)}$University of Sharjah, Sharjah; United Arab Emirates.\\
% ...
$^{103}$CPPM, Aix-Marseille Universit\'e, CNRS/IN2P3, Marseille; France.\\
% ...
$^{122}$Homer L. Dodge Department of Physics and Astronomy, University of Oklahoma, Norman OK; United States of America.\\
% ...
$^{155}$Raymond and Beverly Sackler School of Physics and Astronomy, Tel Aviv University, Tel Aviv; Israel.\\
% ...
$^{a}$ Also at Affiliated with an institute formerly covered by a cooperation agreement with CERN.\\
$^{b}$ Also at An-Najah National University, Nablus; Palestine.\\
$^{c}$ Also at Borough of Manhattan Community College, City University of New York, New York NY; United States of America.\\
$^{d}$ Also at Center for Interdisciplinary Research and Innovation (CIRI-AUTH), Thessaloniki; Greece.\\
$^{e}$ Also at Centre of Physics of the Universities of Minho and Porto (CF-UM-UP); Portugal.\\
$^{f}$ Also at CERN, Geneva; Switzerland.\\
% ...

\end{flushleft}


%

%
%

%

%
%
%
%

\end{document}